\documentclass[11pt,twoside]{article}
\usepackage{asp2006}
\usepackage{epsf}
\usepackage{psfig}
\usepackage{lscape}
\usepackage{graphicx}

\begin{document}
\title{STAR FORMATION HISTORY IN MERGING GALAXIES} 
\author{Li-Hsin Chien} 
\affil{Institute for Astronomy, University of Hawaii} 

\begin{abstract}
Galaxy interactions are known to trigger starbursts.  Young massive star clusters formed in interacting galaxies and mergers may become young globular clusters.  The ages of these clusters can provide clues about the timing of interaction-triggered events, and thus provide an important way to reconstruct the star formation history of merging galaxies.  Numerical simulations of galaxy mergers can implement different star formation rules.  For instance, star formation dependent on gas density or triggered by shocks, predicts significantly different star formation histories.  To test the validity of these models, multi-object spectroscopy was used to map the ages of young star clusters throughout the bodies and tails of a series of galaxy mergers at different stages (Arp~256, NGC~7469, NGC~4676, Arp~299, IC~883 and NGC~2623).  We found that the cumulative distribution of ages becomes shallower as the stage of merger advances.  This result suggests a trend of cluster ages as a function of merger stage.  In NGC~4676 we found that two clusters have ages of about $170$~Myr, suggesting that they likely formed during its first passage.  Their locations in the tidal tails are consistent with the spatial distribution of star formation predicted by shock-induced models.  When comparing the ages and spatial distribution of clusters in NGC~7252 to our model, we found that some clusters are likely to form during the prompt starburst at first passage, as predicted by simulations with shock-induced star formation.  These simulations show that the shock-induced mechanism is an important trigger of star formation and that using the ages of clusters formed in the starbursts can effectively determine the star formation history of merging galaxies.

\end{abstract}

\section{Numerical Simulations of Galaxy Mergers With Star Formation}   

Galaxy collisions provide a natural laboratory for probing how star formation is affected by major rearrangements in the structure and kinematics of galactic disks.  Many observational studies have been devoted to investigating these phenomena and helped to establish the link between galaxy interaction and induced star formation \citep[e.g.][]{kennicutt98,sanders96}.  However, the triggers of star formation in interacting galaxies are still not fully understood.  Studies have suggested two mechanisms to describe the star formation enhancement--- density-dependent \citep[e.g.][]{schmidt59,kennicutt98} and shock-induced \citep[e.g.][]{jog92,scoville86} star formation rules.  Numerical models implementing these rules suggest that simple density-dependent rules cannot offer a complete description of star formation in merging galaxies \citep{mihos93}, and that the two rules predict significantly different star formation histories \citep{barnes04}.  
	
\begin{figure}[!t]
\begin{center}
\includegraphics[width=3.5in]{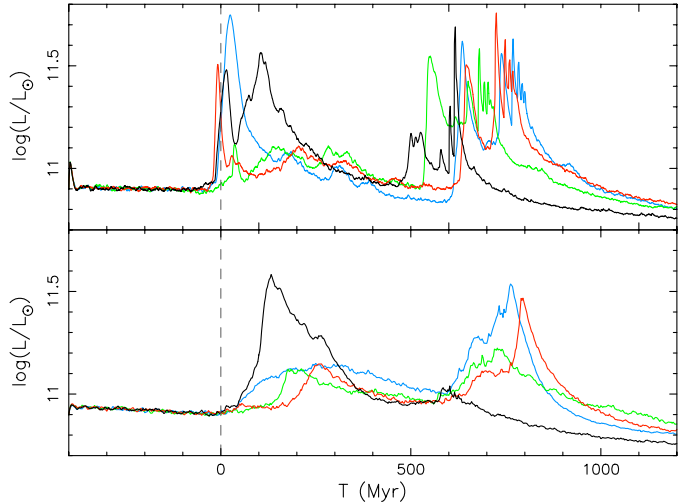}
\caption{Bolometric luminosities as functions of time. Top: shock-induced and Bottom: density-dependent simulations. Colors indicate galaxy pairs colliding with different geometries: Black as DIRECT encounter, red as POLAR, green as INCLINED and blue as RETROGRADE \citep{barnes02}.  Dashed line represents the first passage at T = 0.}

\label{fig1}
\end{center}
\end{figure}
	
Fig.~\ref{fig1} shows bolometric luminosities as functions of time for merger simulations using the two star formation rules and a sample of encounter geometries \citep{barnes02}.  Bolometric luminosity is a good tracer of the star formation rate (SFR) since it comes largely from young massive stars.  Different encounter geometries yield different star formation histories, but the choice of star formation rules is clearly a more important factor.  In shock-induced simulations (top), a global burst is triggered by large-scale shocks during the first passage at $T=0$; later bursts of star formation, concentrated within the central regions, occur with the second passage and merger at $\sim500-600$~Myr.  In contrast, density-dependent models (bottom) generally predict a rather gradual increase within the central regions of the galaxies following the first passage; only the low-inclination passage (RETROGRADE) shows a starburst before the galaxies fall back together and merge.

\section{Young Star Clusters In A Series of Merging Galaxies}   

Violent interactions often trigger starbursts which lead to the formation of young massive star clusters.  These are likely to become young globular clusters (YGCs) if they are still gravitationally bound after $\sim40$~Myr \citep{schweizer99}.  The ages of these YGCs can be interpreted to yield the timing of interaction-triggered events, providing a powerful way to reconstruct the star formation history of merging galaxies.  \citet{chien07} obtained spectra of $12$ young clusters in NGC~4676 using LRIS on Keck.  These spectra yielded reliable approximations for cluster age and metallicity.  Among the ages obtained, two are $\sim170$~Myr, which suggests that they likely formed during the first passage of NGC~4676 \citep{barnes04}.  These two objects are located in the tidal tails of the pair, which is consistent with the spatial distribution of star formation predicted by shock-induced models \citep{barnes04}.

\begin{figure}[thb]
\begin{center}
\includegraphics[width=4in]{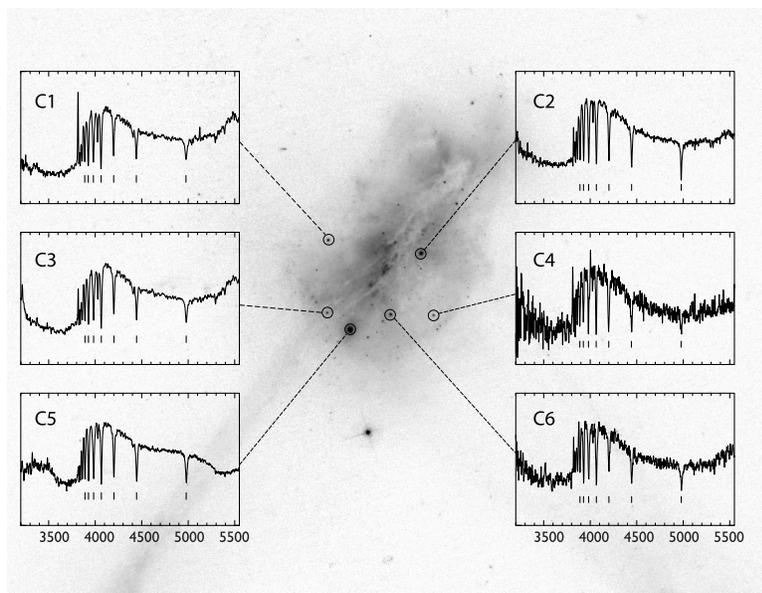}
\caption{ACS/WFC F435W image of IC~883 and spectra of $6$ young clusters obtained with LRIS on Keck. Spectra are plotted as relative flux vs. observed wavelength (\AA). Markers in the spectra are the Balmer series.}
\label{fig2}
\end{center}
\end{figure}

\begin{figure}[thb]
\begin{center}
\includegraphics[width=3in,angle=-90]{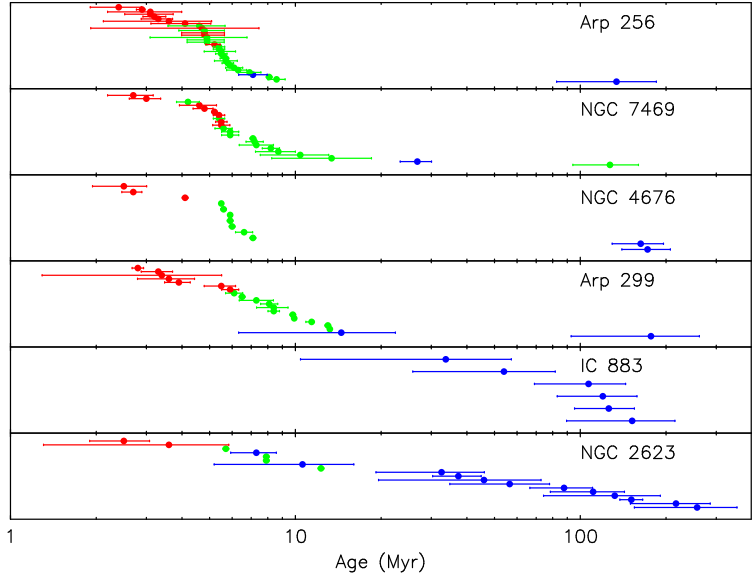}
\caption{Observed cluster age distribution of each galaxy.  For a given panel clusters are plotted according to their age, with the youngest aligned at the top and the oldest at the bottom.  Red points represent clusters with their spectrum dominated by the Balmer emission lines; blue are those dominated by the Balmer absorption lines, and green are those that have composite Balmer features.}

\label{fig3}
\end{center}
\end{figure}

We have also obtained ages of clusters in a series of merging galaxies, ranging from early stages (Arp~256, NGC~7469) through merging (Arp~299) to fully merged (NGC~2623, IC~883) systems.  
For example, Fig.~\ref{fig2} shows spectra of $6$ young clusters in the merged system IC~883.  
Based on our age results, we compare the age distribution of clusters in each galaxy (including NGC~4676) according to their stage of merger (Fig.~\ref{fig3}).  We found more than $70\%$ of the observed clusters have ages less than $10$~Myr in the first two mergers, Arp~256 and NGC~7469, indicating strong on-going star formation in these galaxies.  The ages are distributed more evenly out to about $260$~Myr in the last two merger remnants, IC~883 and NGC~2623, which may suggest that some of these clusters formed during the first or second passages.  This result suggests a trend of cluster ages as a function of merger stages: the cumulative distribution of ages becomes shallower as the stage of mergers advances.  These age distributions provide a crucial way to discriminate between the alternate star formation histories predicted by the two rules described in Sec.~1.  Detailed analysis of ages and metallicities of young star clusters in these galaxies will soon be published \citep{chien09b}.

\section{Combining Observations with Simulations}

\begin{figure}[thb]
\begin{center}
\includegraphics[width=2in,angle=-90]{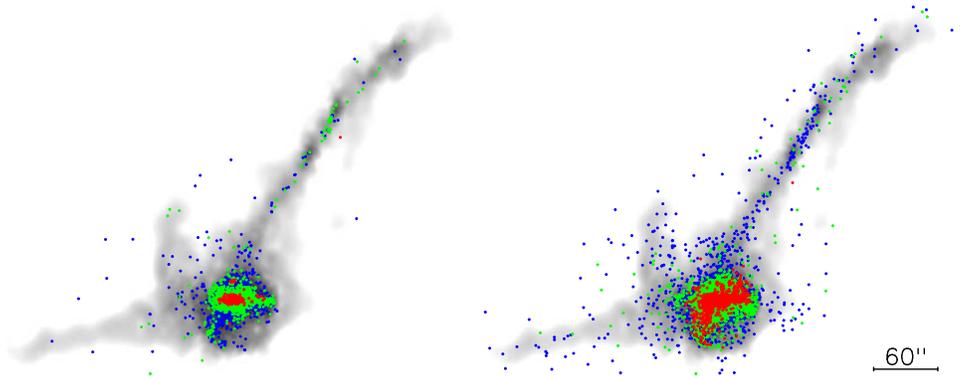}
\caption{Best match of the simulations of NGC~7252 \citep{chien09a}.  Left: density-dependent and Right: shock-induced simulations.  Old stellar particles are shown in grayscale.  Red points are stellar particles with ages $<100$~Myr, green with $400-500$~Myr and blue with $500-600$~Myr. The same number of points are displayed in both images.}

\label{fig4}
\end{center}
\end{figure}

Using interactive software \citep{barnes09} to match dynamical models to the observed morphology and kinematics of mergers, we built new models of NGC~7252 with the two star formation rules described above \citep{chien09a}.  In our models, this proto-elliptical galaxy formed by the merger of two similar gas-rich disk galaxies which fell together $\sim620$~Myr ago.  Fig.~\ref{fig4} shows the spatial distribution of stellar particles of different ages using the two rules.  Although on-going star formation occurs in the central regions in both simulations, the shock-induced simulation predicts that the products of past interaction-induced star formation are also dispersed around the remnant and along the tails.  

\begin{figure}[thb]
\begin{center}
\includegraphics[height=4in,angle=-90]{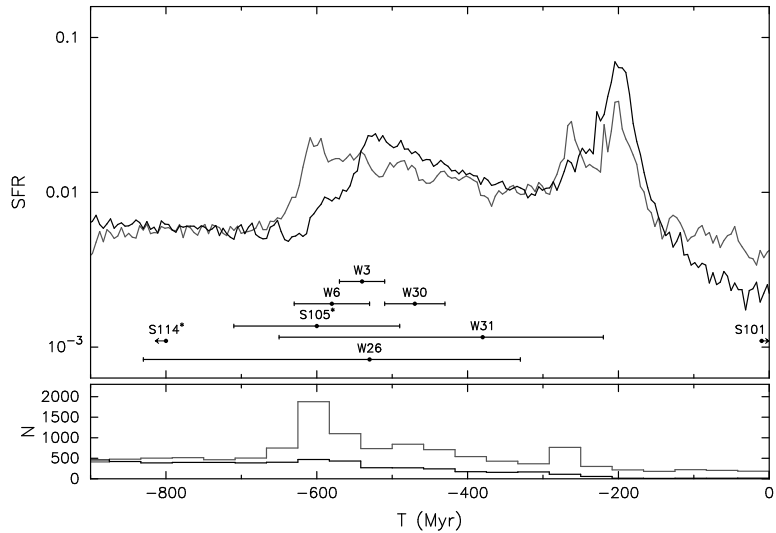}
\caption{Comparison of age distribution of stellar populations.  Top Panel: Global star formation history (in simulation units) of NGC~7252 shown from $900$~Myr ago to present ($T=0$).  Black line represents density-dependent and gray shock-induced simulations.  Cluster ages from \citet{schweizer98} are plotted as dots with their uncertainties.  Cluster S101 has an age upper limit of $10$~Myr and cluster S114 has an age of $\sim1$ Gyr.  Note that clusters S105 and S114 have possible ages of $\sim200$ and $\sim40$~Myr respectively.  Bottom Panel:  Histograms of number of stellar particles formed in the simulation, located within $10\arcsec$ to $35\arcsec$ from the center, measured at present time.}
\label{fig5}
\end{center}
\end{figure}

In addition to comparing our simulations with the observed kinematics and morphologies, we performed a detailed analysis of cluster ages in NGC~7252 (Fig.~\ref{fig5}).  \citet[][hereafter SS98]{schweizer98} found $6$ young clusters which lie between $3-15$~kpc ($\sim10-35\arcsec$) from the center of NGC~7252 and have ages of $\sim400-600$~Myr, indicating that they formed early in the recent merger; the ages of these clusters are plotted in Fig.~\ref{fig5}.  From the top panel we see that the shock-induced simulation (gray) produces prompt burst at first passage $\sim620$~Myr ago while SFR rises more gradually in the density-dependent simulation (black) and peaks $\sim100$~Myr later.  Some of the cluster ages have a wide uncertainty; both of our simulations successfully reproduce the range of cluster ages, although cluster S105 and W6, with ages of $580$ and $600$ Myr, are more consistent with the prompt starburst at first passage in the shock-induced simulation.

However the spatial distribution of the observed clusters strongly discriminates between our models.  To compare with the locations of the SS98 clusters ($10\arcsec<r<35\arcsec$), the bottom panel of Fig.~\ref{fig5} shows the age distribution of star particles, located within this annulus, from our simulations.  The density-dependent simulation (black) produces a gradually declining distribution of ages, with almost all interaction-induced star formation within the very central regions; for example the predicted SFR shows a broad peak around $\sim500$~Myr ago but only a small portion of star particles within this annulus have such ages.  In contrast, the histogram from the shock-induced simulation (gray) shows many more star particles in this annulus and a sharp peak around the first passage $\sim620$~Myr, suggesting that star formation occurs in more dispersed regions away from the centers and that a high portion of star particles within this annulus formed during the starburst in the first passage.  This result may explain the distribution of ages observed in SS98, which indicates that shocks can be an important trigger of formation of these clusters.

In summary, we show that besides the established role of density-dependent mechanism in enhancing the star formation rate in interacting galaxies, the shock-induced mechanism is another important trigger of star formation and that using the ages of clusters formed in the starbursts can effectively pin down the timing of interaction-triggered events and determine the star formation history of merging galaxies.

\acknowledgements
I thank Dr. Josh Barnes, my advisor, for helping me accomplish this research project, and for sharing his great knowledge and insight of interacting galaxies with me.  I am also grateful for helpful discussions with Dr. Francios Schweizer about obtaining ages of clusters and about NGC~7252.  I would like to acknowledge support from the Graduate Student Organization of the University of Hawaii and all the organizers of this conference.



\begin{thebibliography}{}
\bibitem[Barnes(2002)]{barnes02} Barnes, J. E. 2002, MNRAS, 333, 481
\bibitem[Barnes(2004)]{barnes04} Barnes, J. E. 2004, MNRAS, 350, 798
\bibitem[Barnes \& Hibbard(2009)]{barnes09} Barnes, J. E. \& Hibbard, J. E. 2009, AJ, 137, 3071
\bibitem[Chien et al.(2007)]{chien07} Chien, L.-H., Barnes, J. E., Kewley, L. J., \& Chambers, K. C. 2007, ApJ, 660L, 108
\bibitem[Chien \& Barnes submitted(2009)]{chien09a} Chien, L.-H. \& Barnes, J. E. 2009, MNRAS, submitted
\bibitem[Chien et al. in preparation(2009)]{chien09b} Chien, L.-H., Barnes, J. E., Kewley, L. J. \& Evans, A. S. 2009, in preparation
\bibitem[Jog \& Solomon(1992)]{jog92} Jog, C. J. \& Solomon, P. M. 1992, ApJ, 387, 152
\bibitem[Kennicutt(1998)]{kennicutt98} Kennicutt, R. C., Jr. 1998, ARA\&A, 36, 189
\bibitem[Mihos et al.(1993)]{mihos93} Mihos, J. C., Bothun, G. D. \& Richstone, D. O. 1993, ApJ, 418, 82
\bibitem[Sanders \& Mirabel(1996)]{sanders96} Sanders, D. B. \& Mirabel, I. F. 1996, ARA\&A, 34, 749
\bibitem[Schmidt(1959)]{schmidt59} Schmidt, M. 1959, ApJ, 129, 243
\bibitem[Schweizer \& Seitzer(1998)]{schweizer98} Schweizer F. \& Seitzer P. 1998, AJ, 116, 2206
\bibitem[Schweizer(1999)]{schweizer99} Schweizer F. 1999, ASPC, 192, 135
\bibitem[Scoville et al.(1986)]{scoville86} Scoville, N. Z., Sanders, D. B. \& Clemens, D. P. 1986, ApJ, 310, 77
\end{thebibliography}
\end{document}